\documentclass[oldversion]{aa-v6.1}
\usepackage{graphics}

\def\ltsima{$\; \buildrel < \over \sim \;$}
\def\simlt{\lower.5ex\hbox{\ltsima}}
\def\gtsima{$\; \buildrel > \over \sim \;$}
\def\simgt{\lower.5ex\hbox{\gtsima}}

\def\kms{\ifmmode {\rm \ km \ s^{-1}}  
\else
$\rm km \ s^{-1}$\fi}

\def\Msun{\mbox{$M_\odot$}}

\def\sec{\mbox{$^{\prime\prime}$}}
\newcommand{\lessim}{\mathrel{\hbox{\rlap{\hbox{\lower4pt\hbox{$\sim$}}}\hbox{$<$}}}}

\begin{document}
\addtolength{\voffset}{1cm}

%\thesaurus{10.19.3; 08.19.1}

\title{
Mass segregation in rich LMC clusters from modelling of deep HST 
colour-magnitude diagrams}

\author{L. O. Kerber \inst{1,2}
\and B. X. Santiago \inst{1}} 

\offprints{kerber@astro.iag.usp.br}
 
\institute{Universidade Federal do Rio Grande do Sul, IF, 
CP\,15051, Porto Alegre 91501--970, RS, Brazil
\and Universidade de S\~ao Paulo, IAG, Rua do Mat\~ao 1226, 
Cidade Universit\'aria, S\~ao Paulo 05508-900, SP, Brazil}

\date{Received 13 September 2005 / Accepted 10 January 2006}

\titlerunning{Mass segregation in rich LMC clusters}
\authorrunning{L. O. Kerber and B. X. Santiago}

\abstract{
We used the deep colour-magnitude diagrams (CMDs) of
five rich LMC clusters (NGC\,1805, NGC\,1818, NGC\,1831, NGC\,1868, and 
Hodge\,14) observed with HST/WFPC2
to derive their present day mass function (PDMF) and its variation
with position within the cluster. 
The PDMF was parameterized as a power law in the
available main-sequence mass range of each cluster,
typically $0.9 \simlt m/M_{\odot} \simlt  2.5$; its slope 
was determined at different positions spanning from the very centre out to
several core radii. The CMDs in the central regions of the 
clusters were carefully studied earlier, 
resulting in accurate age, metallicity, distance modulus, and reddening values.
The slope $\alpha$ (where Salpeter is 2.35)
was determined in annuli by following two distinct methods:
1) a power law fit to the PDMF obtained from the systemic luminosity 
function (LF);
2) a statistical comparison between observed and model CMDs.
In the second case, $\alpha$ is a free input parameter in the CMD modelling
process  where we incorporate photometric errors and
the effect of binarity as a fraction of unresolved binaries 
($f_{\rm{bin}}=100$ \%) with random pairing of masses from the same 
PDMF.
In all clusters, significant mass segregation is found from 
the positional dependence of the PDMF slope:
$\alpha \simlt 1.8$ for $R \le 1.0$ R$_{\rm{core}}$ and $\alpha \sim$ Salpeter
inside $R=2 \sim 3$ R$_{\rm{core}}$ (except for Hodge 14, where  
$\alpha \sim$ Salpeter for $R \sim 4 $ R$_{\rm{core}}$).
The results are robust in the sense that they hold true for both methods used.
The CMD method reveals that unresolved binaries flatten the PDMF obtained 
form the systemic LF, but this effect is smaller than the uncertainties 
in the $\alpha$ determination.
For each cluster we estimated dynamical ages inside the core and for 
the entire system.
In both cases we found a trend in the sense that older clusters have flatter 
PDMF, consistent with a dynamical mass segregation and 
stellar evaporation.}


\maketitle

\section{Introduction}
\label{intro}
The phenomenon of mass segregation in a stellar system, 
which means a preferential concentration
of high-mass stars towards the centre and a preferential 
allocation of lower-mass stars towards the periphery, seems to occur in 
systems with widely distinct physical properties.
It has been detected and extensively studied in globular clusters 
(de Marchi \& Paresce 1996; Andreuzzi et al. 2000; Howell et al. 2000), 
open clusters (Raboud \& Mermilliod 1998; Durgapal \& Pandey 2001; 
Bonatto \& Bica 2003, 2005), 
and even in star forming regions (Hillenbrand \& Hartmann 1998; 
Stolte et al. 2002).

A key role in the investigation of mass segregation was played by
the Hubble Space Telescope (HST), which for the first time 
resolved the stars in the very centre of the rich star clusters in the 
Magellanic Clouds (MCs). 
This provided a new ``laboratory'' for obtaining constraints on the physical 
processes involved in star formation, where the possible 
universality of the initial mass function (IMF) is a central issue 
(Kroupa 2002). The main reason is
that, unlike the Galaxy, the MCs present a great variety of clusters,
including young and rich star clusters.

There are two possible and distinct origins for this effect of
mass segregation: dynamical and primordial.
The first one is caused by the dynamical evolution of the cluster, where the 
stars tend to reach the equipartition of kinetic energy due to stellar 
encounters (Spitzer 1987; Binney \& Tremaine 1987). 
Therefore, the high-mass stars decrease their velocities, sinking 
towards the cluster centre, while the low-mass stars speed up and
take higher orbits on average.
In a simplified discussion, the characteristic time-scale 
of dynamical mass segregation in a stellar system 
is given by $m_{\rm{low}}/m_{\rm{high}}$ times the two-body relaxation 
time ($t_{rl}$) (Spitzer 1987),  where $m_{\rm{low}}$ and 
$m_{\rm{high}}$ are the lowest and highest masses in the cluster, 
respectively. 
This relation indicates that the time when mass segregation occurs
scales with $t_{rl}$ and can be very short if 
$m_{\rm{high}} >> m_{\rm{low}}$.
An example of a very young ($\simlt 2$ Myr) system 
that presents mass segregation, which can be interpreted as a dynamical 
effect, is the Orion Nebulae Cluster (ONC) (Kroupa, Aarseth \& Hurley 2001).

It is also important to note that dynamical mass segregation 
combined with stellar evaporation may lead to the preferential loss 
of lower-mass stars.
Since these stars are more likely populating the outermost regions of 
a cluster, they are more prone to be unbound due to their lower
binding energy.
Therefore one could expect that stellar clusters would have flatter 
global PDMF as they become dynamically older.
In fact this effect is observed in open clusters (Bonatto \& Bica 2005)
and Galactic globular clusters (Baumgardt \& Makino 2003), which are
also modelled with N-body simulations.

On the other hand, primordial mass segregation may be a natural
outcome of star-formation theory, since a protocluster with higher 
central density should have a greater probability of forming proportionally 
more high-mass stars in its centre. Some scenarios propose mass
segregation at the onset of star formation through
interactions among the protostars, since the
collision probabilities increase 
with density (Bonnel \& Davies 1998); alternatively, accretion rates are
enhanced with the mass of the accreting protostar (Behrend \& Meader 2001). 

Regardless of the physical mechanism used to account for the 
origin of mass segregation, there are several techniques for
diagnosing and quantifying the effect, the main ones being based on 
stellar statistics.
By counting stars in different annuli, one may search for variations
in radial profiles as a function of stellar-mass range or for
changes in the slope of the luminosity function (LF) or of 
the present day mass function (PDMF). 
Although there are uncertainties in the conversion of 
stellar luminosity into mass (de Grijs et al. 2002a), 
the second option provides a more direct
constraint on N-body simulations that intend to recover the initial conditions 
of the cluster (Kroupa, Aarseth \& Hurley 2001; 
Baumgardt \& Makino 2003; Moraux, Kroupa \& Bouvier 2004 ).
Since the mass-luminosity relation is dependent on metallicity, its 
uncertainty may be efficiently reduced by precisely determining the
cluster physical parameters, by means of careful 
modelling of its colour-magnitude diagram (CMD).
A detailed enough CMD modelling should result in physical parameters 
predominantly limited by uncertainties associated to the models of 
stellar evolution, rather than to the data.

The main goal of this work is to determine the spatial dependence of 
the PDMF slope of five rich Large Magellanic Cloud (LMC) clusters, 
namely NGC\,1805, 
NGC\,1818, NGC\,1831, NGC\,1868, and Hodge\,14.
In Kerber \& Santiago (2005), we presented the analysis of deep CMDs
from these clusters 
obtained with HST/WFPC2 in the {\it F555W} ($\sim V$) and {\it F814W} 
($\sim I$) filters. 
Efficient use of the data was made by means of direct comparisons
of the observed CMD (statistically corrected for incompleteness and field 
star contamination) to model ones. 
By modelling the CMDs in the central region of each cluster we inferred the
metallicity ($Z$), the intrinsic distance modulus ($(m-M)_{0}$) and the 
reddening value ($E(B-V)$). We also determined the age ($\tau$)
for NGC\,1831, NGC\,1868, and Hodge\,14

Santiago et al. (2001) and de Grijs et al. (2002ab), using the same data, 
analyse the mass segregation in these clusters by means of the spatial 
dependence of the LF slope.
Their diagnostic was clear: all clusters present mass segregation, even
the youngest ones (NGC 1805 and NGC 1818). 
Meanwhile, de Grijs et al. (2002ab) reach the same result by comparing 
the radial profile dependence with the stellar mass range. 
They also derive the PDMF and its variation with position within the 
cluster, but only for the youngest ones. 
Here we present this type of analysis for the five clusters, using the 
cluster parameters derived by Kerber \& Santiago (2005).
The PDMF slope ($\alpha$) in different annuli was determined by following 
two distinct methods: 
1) power law fit to the PDMF obtained from the systemic LF 
(hereafter LF method);
and 2) a statistical comparison (similar to Kerber \& Santiago 2005) 
between observed and model CMDs 
(hereafter CMD method), where $\alpha$ is a free input 
parameter in the CMD modelling process.
The main difference between the two methods is that the CMD method 
potentially uses all the information contained in the CMD, including 
the effects of unresolved binaries and photometric uncertainties.

The paper is divided as follows.
In Sect. 2 we present a brief description of our data and the physical 
properties of the clusters. 
In Sect. 3 we present the two methods of determining the PDMF slope and
their results. 
These results are then discussed in Sect. 4, where we also compare them 
with those available in the literature.
Finally, in Sect. 5 we present a summary and our concluding remarks.

\section{The data}
\label{data}

We used data taken with HST/WFPC2 as part of a cycle 7 project entitled
``Formation and evolution of rich LMC clusters'' (Beaulieu et al. 1999). 
For each cluster and a nearby field, images were obtained using the 
{\it F555W} ($\sim V$) and {\it F814W} ($\sim I$) broad band filters. 
A detailed description of the photometry and sample completeness corrections 
can be found in Santiago et al. (2001) and Castro et al. (2001). 
Kerber \& Santiago (2005) made a detailed study of the resulting CMDs 
in the central cluster regions, in order to infer the cluster's global 
parameters, such as age, metallicity, foreground extinction,
and distance. Their work was based on a detailed CMD modelling process.
In brief, the modelling is based on
the generation of synthetic CMDs to be compared to the observed one. 
For a realistic comparison to the models, the observed CMDs had to be 
corrected for selection
effects, such as photometric incompleteness and field-star contamination. 
Random photometric uncertainties in the real data 
were also quantified and applied to the model CMDs.
Finally, different statistical tools were used in the model vs. data 
CMD comparison.

Table \ref{tab1} lists the main parameters of the clusters 
in the sample.
The cluster name is followed by the total number of stars 
($N_{\rm{clus}}$) in its final CMD after being corrected for the 
effects mentioned earlier (see Kerber \& Santiago 2005 for details).
This table also lists the cluster's adopted core radius ($R_{\rm{core}}$, 
as determined by Mackey \& Gilmore 2003);
and the radius at which the star density falls to the field
density ($R_{\rm{max}}$).
The metallicity ($Z$), logarithmic age ($\tau$),
intrinsic distance modulus ($(m-M)_{0}$), and reddening value ($E(B-V)$)
were kept fixed in the PDMF analysis and are consistent with the ones 
determined by Kerber \& Santiago (2005).

\begin{table}
\caption[]{Main parameters of the clusters in the sample.}
\label{tab1}
%\small
\renewcommand{\tabcolsep}{0.3mm}
\begin{tabular}{lccccccc}
\hline\hline
Cluster & $N_{\rm{clus}}$ & $R_{\rm{core}}$ & $R_{\rm{max}}$ & Z & log($\tau$/yr) & $(m-M)_{0}$ & $E(B-V)$ \\
~~ & ~~ & ($\sec$) & ($\sec$) & ~~ & ~~ & ~~ & ~~ \cr
\hline
NGC\,1805 & 2564 & 5.5 & 55 & 0.008 & 7.80${^*}$ & 18.55 & 0.03 \cr
NGC\,1818 & 3929 & 10.1 & 60 & 0.004 & 7.80${^*}$ & 18.45 & 0.00 \cr
NGC\,1831 & 7136 & 18.3 & 90 & 0.012 & 8.70 & 18.70 & 0.00 \cr
NGC\,1868 & 5675 & 6.7 & 70 & 0.008 & 8.95 & 18.70 & 0.00 \cr
Hodge\,14 & 1196 & 7.4 & 40 & 0.006 & 9.25 & 18.55 & 0.03 \cr
\hline
\end{tabular}

$^{*}$youngest isochrone available by Girardi et al. (2000).
\end{table}

\section{Analysis}

The PDMF is considered here to be a power law:

\begin{equation}
\label{eq1}
\xi(m) =\frac{{\rm{d}}N}{{\rm{d}}m} = \xi_{0}~m^{-\alpha}
\end{equation}

\noindent
where $\xi_0$ is a normalization constant and $\alpha$ the PDMF slope.
Different parameterizations are found in the literature, often making use
of multiple power laws. Given the limited mass range of main-sequence
stars in the CMDs studied here, use of a single slope is justified.

To analyse the positional dependence of $\alpha$ within a cluster,
we divided its CMD stars into subsamples, 
according to the distance from the cluster centre, defining 
several annuli.
The inner and outer radial limits of each ring were chosen to 
ensure that each subsample would typically contain 600 stars, 
allowing a statistically significant number of stars to contribute
to each PDMF determination.  

To determine the PDMF slope ($\alpha$) in each ring we used two 
distinct approaches, each one made up of several steps:

\noindent
{\bf 1)} derivation of the systemic LF; conversion of the LF
into a PDMF, according to a given mass-luminosity relation;
a power law fit to the PDMF obtained from the 
systemic LF We refer to this method as the LF method. 

\smallskip

\noindent
{\bf 2)} generation of synthetic CMDs with different input PDMFs,
keeping the binary fraction and model isochrone parameters fixed; 
statistical comparison between observed and model CMDs; determination
of the best-fitting CMD models for each data CMD.
In this case, the PDMF slope $\alpha$ is a free input parameter of the 
CMD models. As the statistical comparison in this case makes use of the 
information available in the entire CMD plane, we call it the CMD method.

Both approaches require conversion from mass to luminosity (CMD method) 
or vice-versa (LF method).
The mass-luminosity relation used in these conversions is provided by 
Padova isochrones (Girardi et al. 2000) that are
shifted in magnitude and colour by $(m-M)_{0}$ and $E(B-V)$ and 
where the parameters (including $Z$ and $\tau$) are as given in 
Table \ref{tab1}. 
When necessary, we interpolated the original Padova isochrone 
grid in metallicity (see Kerber \& Santiago 2005, Sect. 3.2 and their Fig. 13)
in order to generate an isochrone with the quoted values.

Each approach displays its own advantages and disadvantages.
The LF method, by construction, does not use all the two-dimensional
information contained in the CMD plane. 
This may actually be considered an advantage, since the
PDMF can be reliably recovered with a smaller number of stars, as
each magnitude bin concentrates the information spread along the MS colour
width.
On the other hand, the effect caused by unresolved binaries may
be crucial, since the CMD position of primary stars will be spread
redwards and brightwards due to the presence of the secondaries. 
If unaccounted for, unresolved binarity will cause some shallowing 
in the recovered PDMF, as the system masses resulting from their 
combined luminosities will be larger than the masses of the 
individual components.

Several previous works have taken the effect of unresolved 
binaries into account in the recovered PDMF or IMF from a systemic LF.
Kroupa, Tout, \& Gilmore (1991) have managed to reconcile the LF of local 
volume-limited samples of Galactic field stars with the LF inferred from 
photometric surveys of more distant by correcting them for the 
systematics caused by unresolved pairs.
More recently, Kroupa (2001) showed that the single-star IMFs 
can be systematically steeper by 0.5 between $0.1 \le m \le 1.0~M_{\odot}$
than the Galactic-field IMF. 
Studying LMC star clusters, Sagar \& Richtler (1991) determined that
the recovered PDMF from the systemic LF can become significantly 
steeper (by $\Delta \alpha \simeq 1.0$) if the binary fraction is 
large ($f_{\rm{bin}} \simgt 0.50$) and $\alpha \sim 0.5$.
Similar techniques were applied by Sandhu, Pandey, \& Sagar (2003) to
correct the PDMF for intermediate/old open clusters.

In order to simulate the effect of unresolved binaries in PDMF 
determination we introduced it in the CMD modelling process. The well-known 
signature of unresolved pairs in a CMD (see Hurley \& Tout 1998 for a  
demonstration of the effect) can be modelled in a straightforward manner by
 applying the combined fluxes and colours from two stars 
to a given fraction ($f_{\rm{bin}}$) of systems. 
One caveat may be the uncertainties in the
distribution ($\rm{d}N/\rm{d}q$) of secondary/primary mass ratios 
($q=m_{2}/m_{1} \le 1.0$, 
where $m_{1}$ and $m_{2}$ are, respectively, 
the primary and secondary masses).
This may at first sight be regarded as an extra degree of freedom in the
modelling process, since secondary star masses may not necessarily
be drawn from the same distribution as the primary stars.
However, this possibility poses a question of what a mass function is 
meant to be, as it would not be uniquely defined even in a single population. 
Here we adopt the assumption that secondary stars in binary systems have 
masses drawn from the same PDMF as primary stars or as single stars. 
To maximize the possible effect of unresolved binaries and therefore
explore the most of this effect, we used $f_{\rm{bin}}$=100\%.

Furthermore, the CMD modelling naturally incorporates the photometric
uncertainties into the PDMF determination, and can potentially incorporate 
other observational effects that may influence the conversion of magnitude 
and colour information into mass. 
The major disadvantage of a CMD method is the prior 
PDMF parameterization, here modelled as a power law with one free-parameter 
for $m \ge 0.80~M_{\odot}$ ($\simgt$ observed lower-mass limit).
Even though it does not directly affect the PDMF in the observed regime, 
we fixed a shallow slope $\alpha=1.30$ 
in the lower mass range ($0.08 \le m/M_{\odot} \le 0.80$). 
This is consistent with the IMF proposed by Kroupa (2002) 
in this mass range. 
This relatively shallow slope in the low-mass range
also yields an enhanced effect of unresolved binaries.
Although in our simulations all stars have one companion, 
only pairs with $q \simgt 0.60$  will significantly change the CMD 
position of the primary star.
The fraction of such effective binary systems, $f_{\rm{bin,eff}}$,
is $\sim 20\%$. They are practically the only ones responsible for the 
effect that unresolved binaries may cause in the CMD or in the LF.
We refer to Tout (1991) for a demonstration of the sensitivity
of PDMF slope with the mass-ratio distribution in the low-mass
regime.

\subsection{LF method}

\begin{figure}
\resizebox{\hsize}{!}{\includegraphics{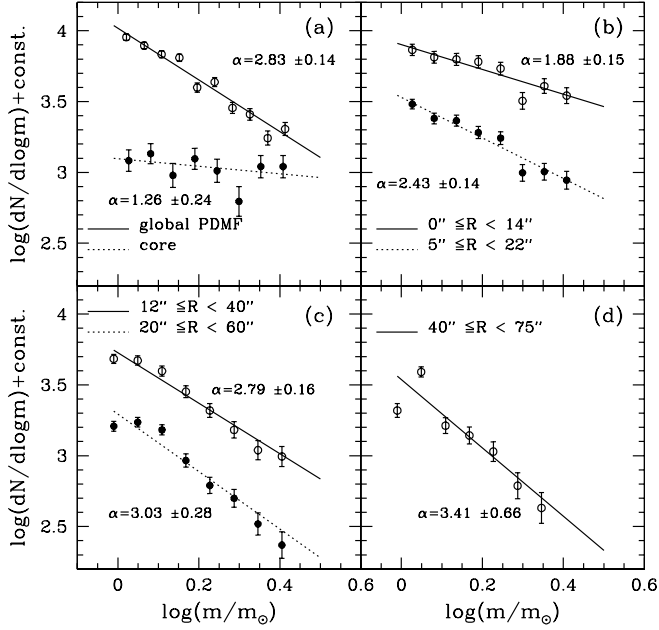}}
\caption[]{PDMFs at different concentric regions around NGC 1805.
The lines are power law fits to the PDMFs. The slope values and their
corresponding fit errors are given
in each case, as well as the inner and outer radii of each region.}
\label{n1805_pdmf}
\end{figure}

\begin{figure}
\resizebox{\hsize}{!}{\includegraphics{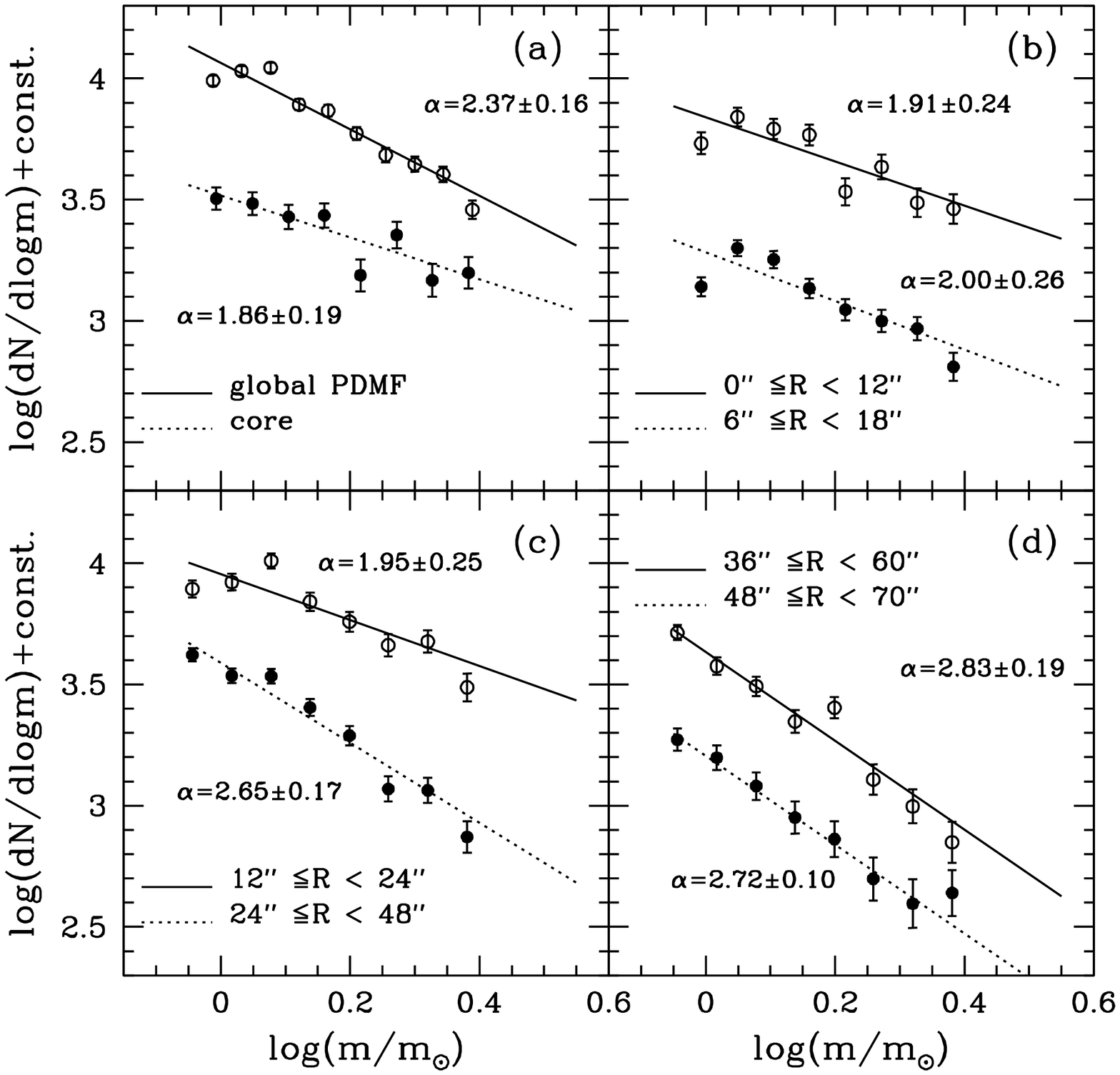}}
\caption[]{Same as in Fig. \ref{n1805_pdmf}, but now for NGC 1818}
\label{n1818_pdmf}
\end{figure}

\begin{figure}
\resizebox{\hsize}{!}{\includegraphics{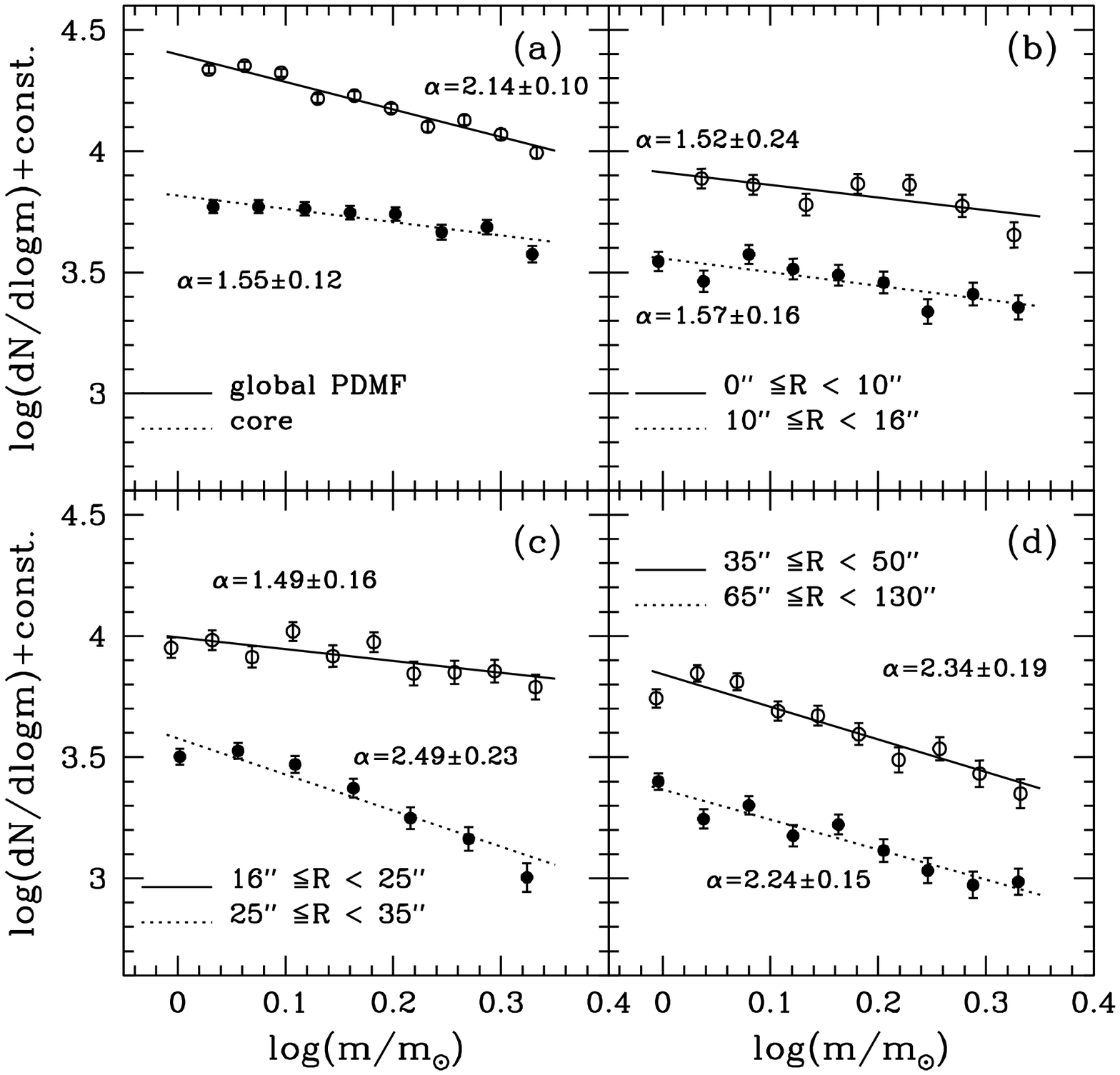}}
\caption[]{Same as in Fig. \ref{n1805_pdmf}, but now for NGC 1831}
\label{n1831_pdmf}
\end{figure}

\begin{figure}
\resizebox{\hsize}{!}{\includegraphics{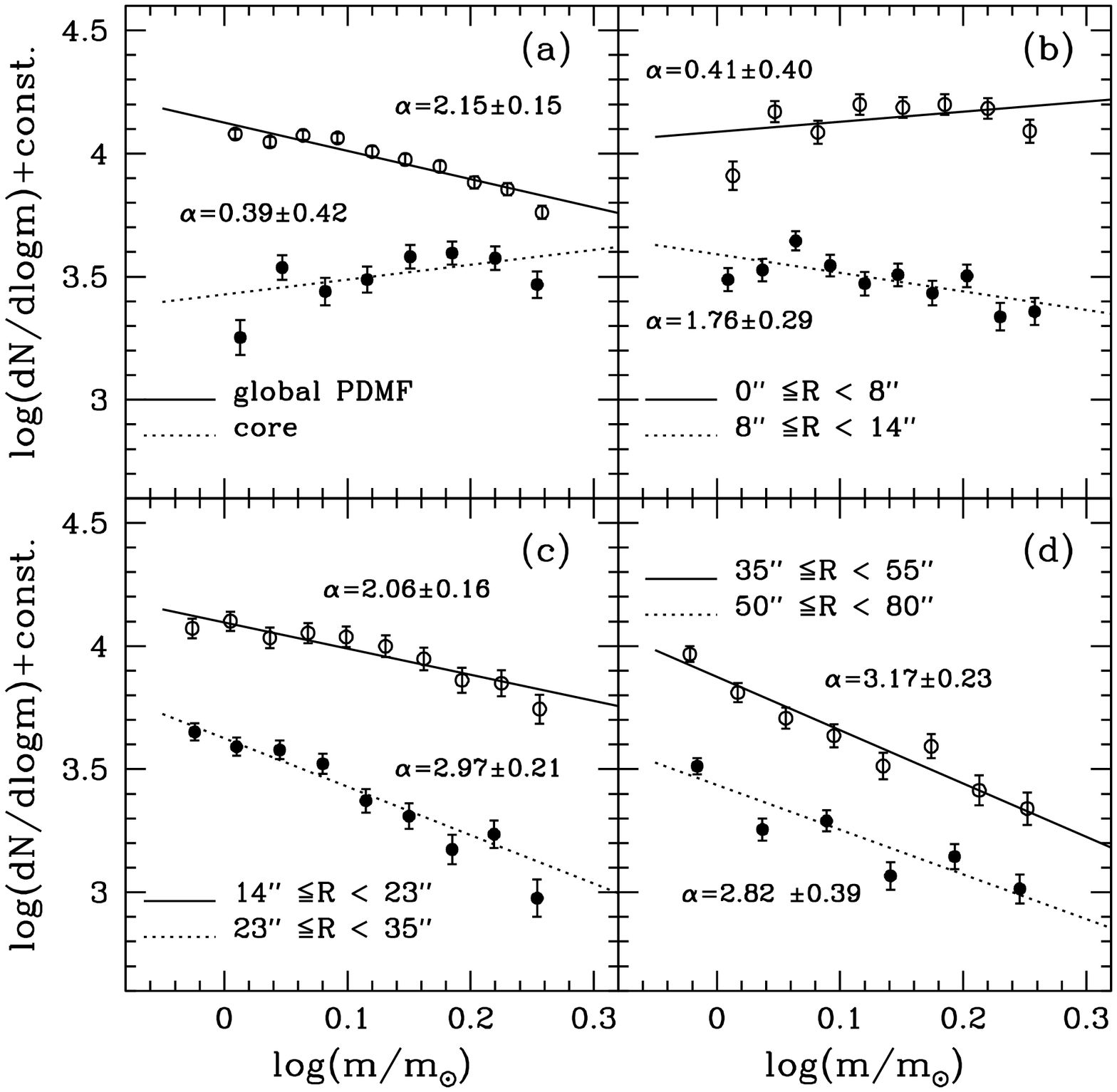}}
\caption[]{Same as in Fig. \ref{n1805_pdmf}, but now for NGC 1868}
\label{n1868_pdmf}
\end{figure}

\begin{figure}
\resizebox{\hsize}{!}{\includegraphics{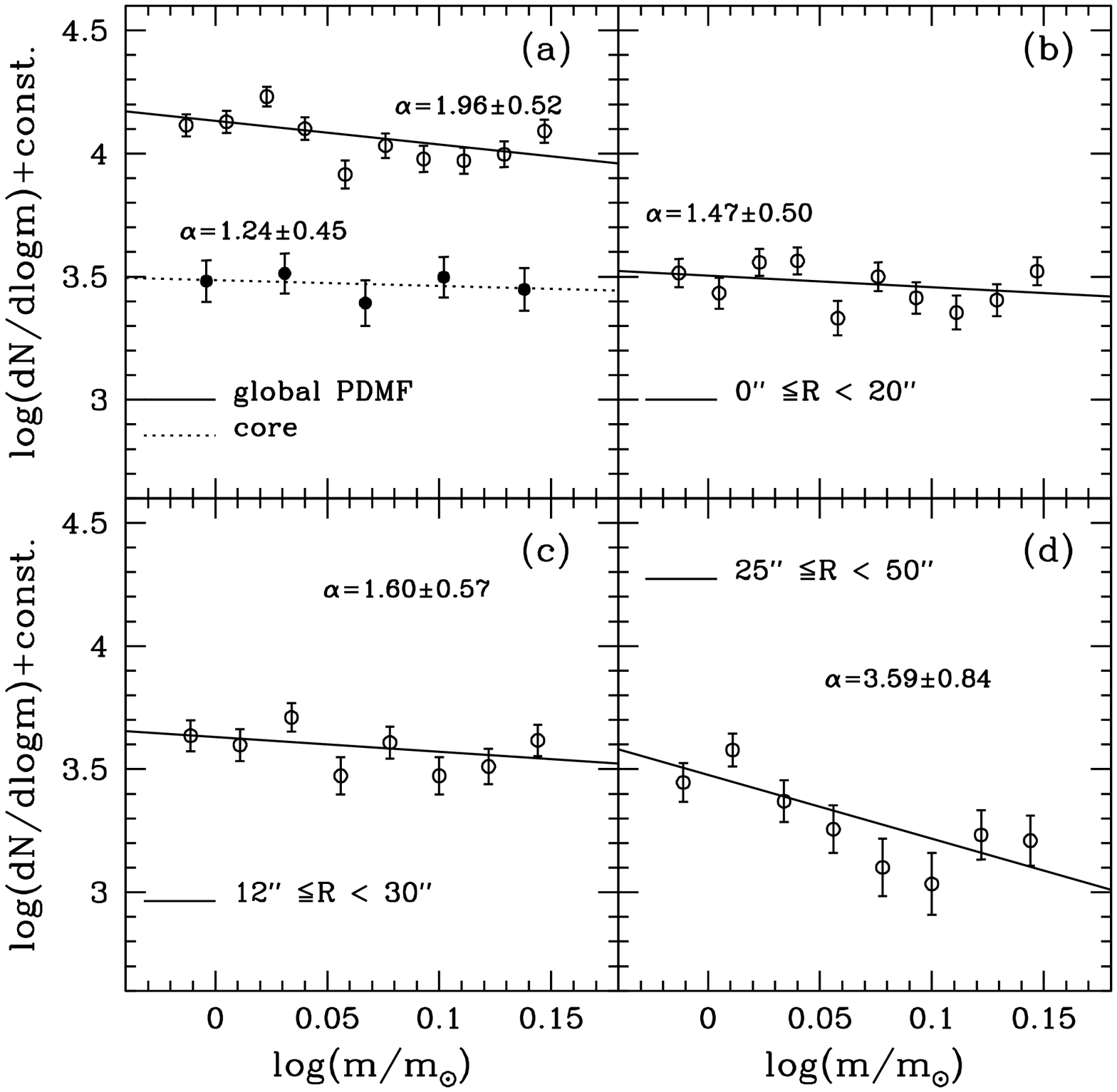}}
\caption[]{Same as in Fig. \ref{n1805_pdmf}, but now for Hodge 14}
\label{h14_pdmf}
\end{figure}

Mass segregation was observed by Santiago et al. (2001)
in their study of the LFs of the clusters in our sample.
Then de Grijs et al. (2002a) converted LFs into PDMFs and presented 
their positional dependence for the younger clusters, NGC 1805 and NGC 1818.
But those authors
did not use CMD modelling techniques to investigate the behaviour 
of $\alpha$ as a function of distance from the cluster centre. 
Also, the mass-luminosity relations used in previous works were not
the same as those used here, which are based on careful constraints on the
cluster's ages and metallicities, which come from our previous CMD modelling.

Notice that the mass-luminosity relation is a key ingredient in 
conversion from an LF into a PDMF, a traditional procedure 
widely used in stellar statistics.
By definition, the LF ($\Phi(M_{555})$) in the $M_{555}$ 
absolute magnitude is given by 

\begin{equation}
\label{eq2}
\Phi(M_{555}) = \frac{{\rm{d}}N}{{\rm{d}}M_{555}}~,
\end{equation}

\noindent
where ${\rm{d}}N$ is the number of individual stars that have absolute 
magnitude inside $(M_{555}, M_{555}+{\rm{d}}M_{555})$.

If the PDMF is expressed in linear mass bins ($\xi(m)$),
it is related to the LF as follows 

\begin{equation}
\label{eq3}
\xi(m)=\frac{{\rm{d}}N}{{\rm{d}}m} = 
- \Phi(M_{555})~\left[\frac{{\rm{d}}m(M_{555})}{{\rm{d}}M_{555}}\right]^{-1}
\end{equation}

\noindent
where the $m(M_{555})$ is the mass-luminosity relation.  
Here one clearly sees the importance of the choice of 
suitable stellar evolution models, 
as they provide not only the mass-luminosity relation, but also 
its derivative. As previously discussed, 
it is important to keep in mind that the observed photometric
data suffer from unresolved binarity and therefore allow us
to construct only the systemic LF.
As a consequence, there is not a unique mapping from one $M_{555}$ 
magnitude to one stellar mass; the LF method may be seen as a first 
attempt to recover the PDMF. 
We refer the reader to Sect. \ref{bineffect}, 
where we present an approach to evaluating the 
effect of unresolved pairs in the recovered PDMF from a systemic LF.

Analogous to the LF, the PDMF may also be expressed 
in logarithmic ($\rm{log_{10}}$)
mass bins ($\xi{_{\rm{L}}}(m)$), in which case
it is related to $\xi(m)$ by

\begin{equation}
\label{eq3}
\xi{_{\rm{L}}}(m)=\frac{{\rm{d}}N}{{\rm{dlog}}m} = 
\xi(m)~\left[\frac{{\rm{dlog}}m}{{\rm{d}}m}\right]^{-1}=
{\rm{log}}e~m~\xi(m).
\end{equation}

\noindent
Also parameterizing the $logarithmic$ PDMF as a power law, 
we have $\xi_{{\rm{L}}}(m)=\xi_{{\rm{L}},0}~m^{\Gamma}$, 
whose slope $\Gamma$ is correlated with $\alpha$ by  
$\Gamma = -(\alpha-1)$.

Taking the logarithm on both sides of Eq. \ref{eq1}, we have
${\rm{log}}\xi = -\alpha{\rm{log}}m + {\rm{log}\xi_{0}}$ or 
$ {\rm{log}}\xi_{{\rm{L}}} = -(\alpha-1){\rm{log}}m + {\rm{log}\xi_{0}}
-0.362$. In other words, in a log-log plot, both the linear and logarithmic 
power law PDMFs should behave as a straight line, whose slope yields $\alpha$
directly.

\begin{figure*}
\resizebox{\hsize}{!}{\includegraphics{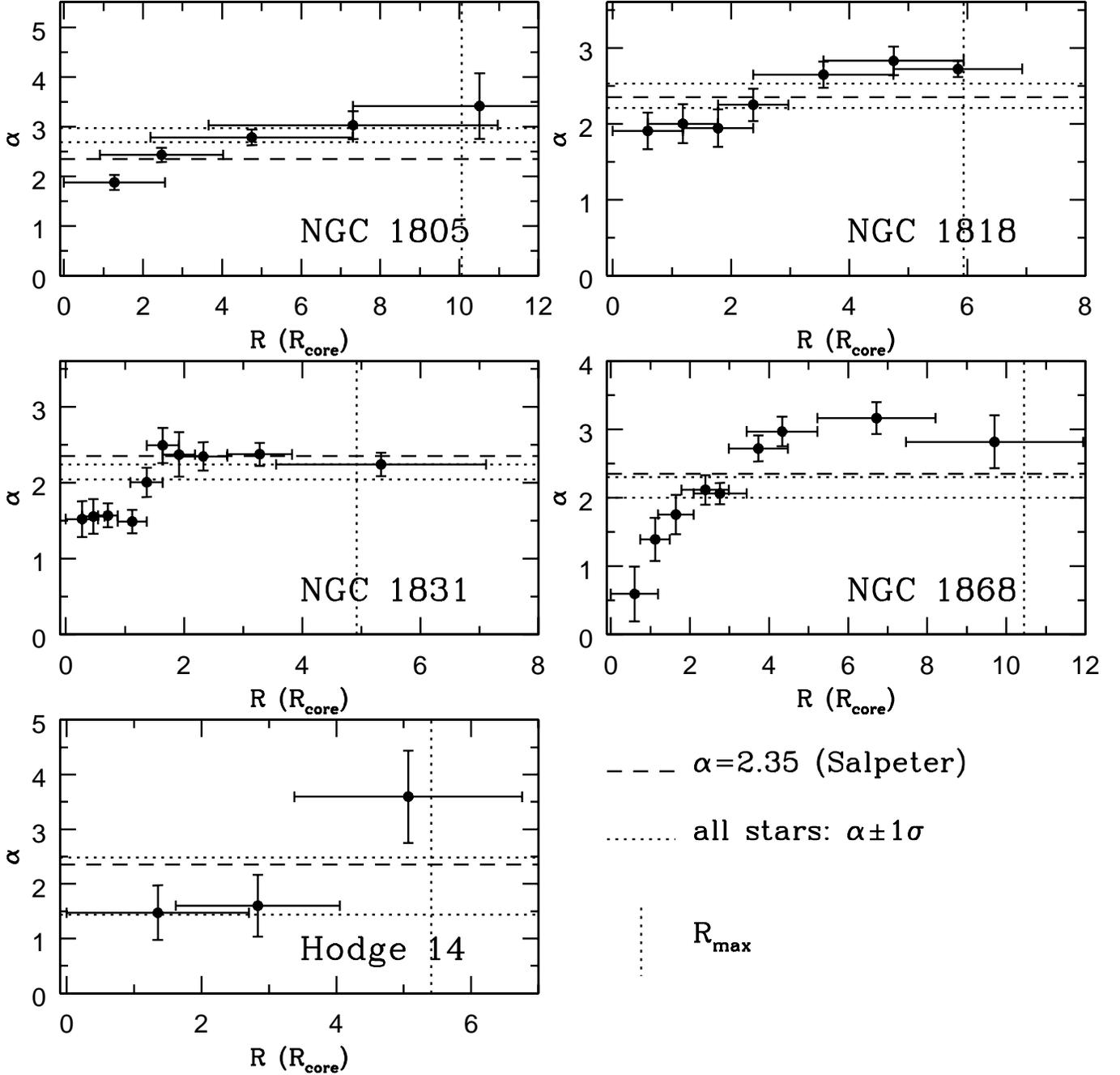}}
\caption[]{The PDMF slope {\it vs} R inferred from the LF analysis. 
The horizontal dotted lines show the 2$\sigma$ range in 
the global PDMF slope.
The horizontal dashed line is the reference Salpeter value ($\alpha$=2.35).
The $R_{\rm{max}}$ is marked as the vertical line.}
\label{alpha_R_1D}
\end{figure*}

Figures {\ref{n1805_pdmf}}-{\ref{h14_pdmf}} present the behaviour of the
PDMFs recovered directly (without any treatment of the unresolved 
binarity) from the systemic LFs at different 
concentric regions around each cluster.
For convenience, we determine $\xi_{\rm{L}}$ and plot it in a 
logarithmic scale, where we shift 
the values of $\rm{log}(\xi_{\rm{L}})$
up or down to avoid overlapping in the panels.
In each case $\alpha$ was obtained by means of a linear fit,
shown as a straight line in the figure panels. 
Panel a) shows the PDMF for all stars in each cluster and the PDMF
for the stars within the core.
In all cases $\alpha$ is smaller in the core than for the entire cluster, 
which indicates mass segregation. 
Besides, these figures show the expected signature of mass segregation:
the PDMF is shallower (smaller $\alpha$) in the inner regions (panel b)
than in the outer regions (panel d). 
Also, in general, the single slope fit is a very adequate description of the
data.
The exceptions to this rule often result from a sudden drop 
at the low-mass end (log($m/M_{\odot}$) $\sim$ 0.0, $m \sim 1.0~M_{\odot}$)
of the PDMFs in the central regions of
the richest clusters, such as NGC 1868. 
These deviations from a power law are very likely due to residuals in the 
correcting for incompleteness effects.

Figure \ref{alpha_R_1D} plots $\alpha$ as a function of distance
$R$ from the cluster centre. This distance is expressed in units
of the core radius, listed in Table  \ref{tab1}. 
The effect of mass segregation is again clearly seen for all clusters.
The vertical bars on each point
represent the linear-fit uncertainties in $\alpha$. 
The horizontal bars just show the limits of the annuli.
Some concentric regions are partially overlapping in order 
to yield a more continuous behaviour of the PDMF slope 
and to increase the statistical significance in each PDMF determination. 
The dotted-horizontal lines constrain the $1\sigma$ range of
$\alpha$ fitted to the global PDMF.
This range can be compared with the Salpeter (1955) value 
($\alpha=2.35$), marked as a dashed-horizontal line.
In all the others clusters, besides Hodge 14, $\alpha(R)$ has a
Salpeter value around $R \simeq 2-3$ R$_{\rm{core}}$. In the outer
regions, the relation $\alpha$ vs. $R(R_{\rm{core}})$ flattens for NGC 1818,
NGC 1831, and NGC 1868. This may be the result of the dynamical 
loss of lower-mass stars from the clusters.

\subsection{CMD method}

Only the systemic LF, 
which suffers from the effect unresolved binaries, is directly 
extracted from the data, as discussed in the 
introduction to this section. Therefore, we deal in this section with
the 2D CMD modelling process, an approach that is capable of explicitly
taking unresolved binarity into account.

Our CMD modelling process and the statistical techniques of CMD comparisons
were extensively explained in Kerber \& Santiago (2005) and 
Kerber et al. (2002). Here we only underline some important aspects. 
The modelling process assumes that the cluster is a single stellar population
(SSP) that generates a synthetic main-sequence (MS) in the CMD plane, 
where we introduce as model inputs the information about metallicity ($Z$), 
age ($\tau$) (given by a Padova isochrone; Girardi et al. 2000), 
intrinsic distance modulus ($(m-M)_{0}$), reddening value ($E(B-V)$), 
PDMF slope ($\alpha$), and fraction of unresolved binaries ($f_{\rm{bin}}$).
By exploring a regular model grid, we may then find the best models by 
means of statistical comparisons carried out in 1 and 2 dimensions and, 
therefore, the physical parameters that best constrain the cluster CMD. 
By modelling the CMDs in the central region, Kerber \& Santiago (2005) 
inferred $Z$, $(m-M)_{0}$, and $E(B-V)$ for each cluster. 
They also determined the age of three clusters: NGC 1831, 
NGC 1868, and Hodge 14. 
A set of values consistent with these determination is 
listed in Table \ref{tab1}.

\begin{figure}
\resizebox{\hsize}{!}{\includegraphics{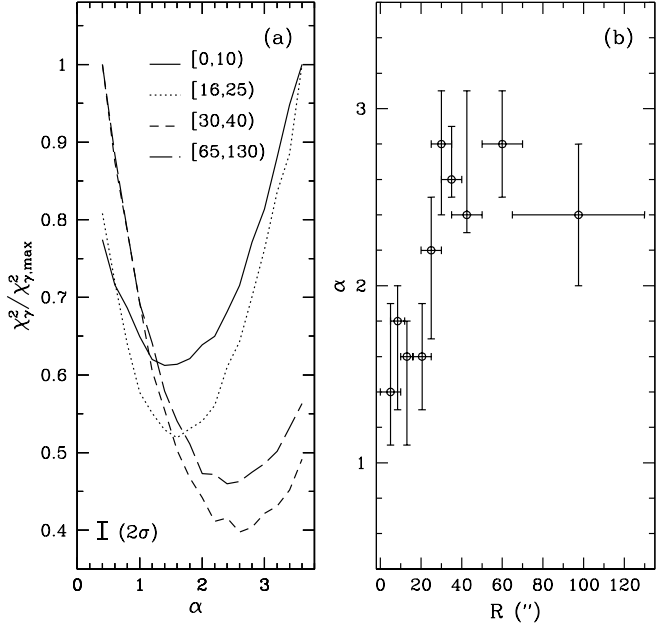}}
\caption[]{The CMD method for NGC 1831. Panel {\bf a} illustrates
how the statistic $\chi^{2}_{\gamma}$ varies with the adopted 
$\alpha$ used in the CMD models at different concentric 
annuli. Panel {\bf b} shows the best-fit $\alpha$ {\it vs} R relation.}
\label{distchi2}
\end{figure}

We here model the CMDs separated by annuli
 in order to investigate the positional dependence of the only 
remaining free parameter: $\alpha$.
Figure \ref{distchi2} illustrates the CMD method for NGC 1831. 
Panel a) shows how the statistic $\chi^{2}_{\gamma}$ varies
as a function of the input value of $\alpha$ used to build the synthetic CMDs.
Each curve represents the run of $\chi^2_{\gamma}$ (normalised
by its maximum value, $\chi^2_{\gamma,max}$) with $\alpha$ at
a given cluster ring. The effect of mass segregation is already clear
in this figure, as the minimum value $\chi^{2}_{\gamma,min}$ occurs
at higher $\alpha$ as the outer annuli are considered, reflecting
a gradual steepening in the PDMF. 

In order to determine $\alpha$ and its associated uncertainty from this method,
100 realisations of the model that yielded $\chi^{2}_{\gamma,min}$ were
run, from which a dispersion in $\chi^2_{\gamma}$ for this model, 
$\sigma_{\chi}$, was determined. Therefore, the models that satisfy 
the criterion

$$\chi^{2}_{\gamma} \leq \chi^{2}_{\gamma,min} + \sigma_{\chi}$$

\noindent
are statistically of similar quality as the one that yields the minimum.
The average value of $\alpha$ for these models 
is then considered as the one that best describes the PDMF, and the standard
deviation around this average represents its uncertainty. 
Finally, panel b) in Fig. \ref{distchi2} shows the $\alpha$ values 
determined for all annuli, 
again revealing variations with position inside the cluster.

\begin{figure*}
\resizebox{\hsize}{!}{\includegraphics{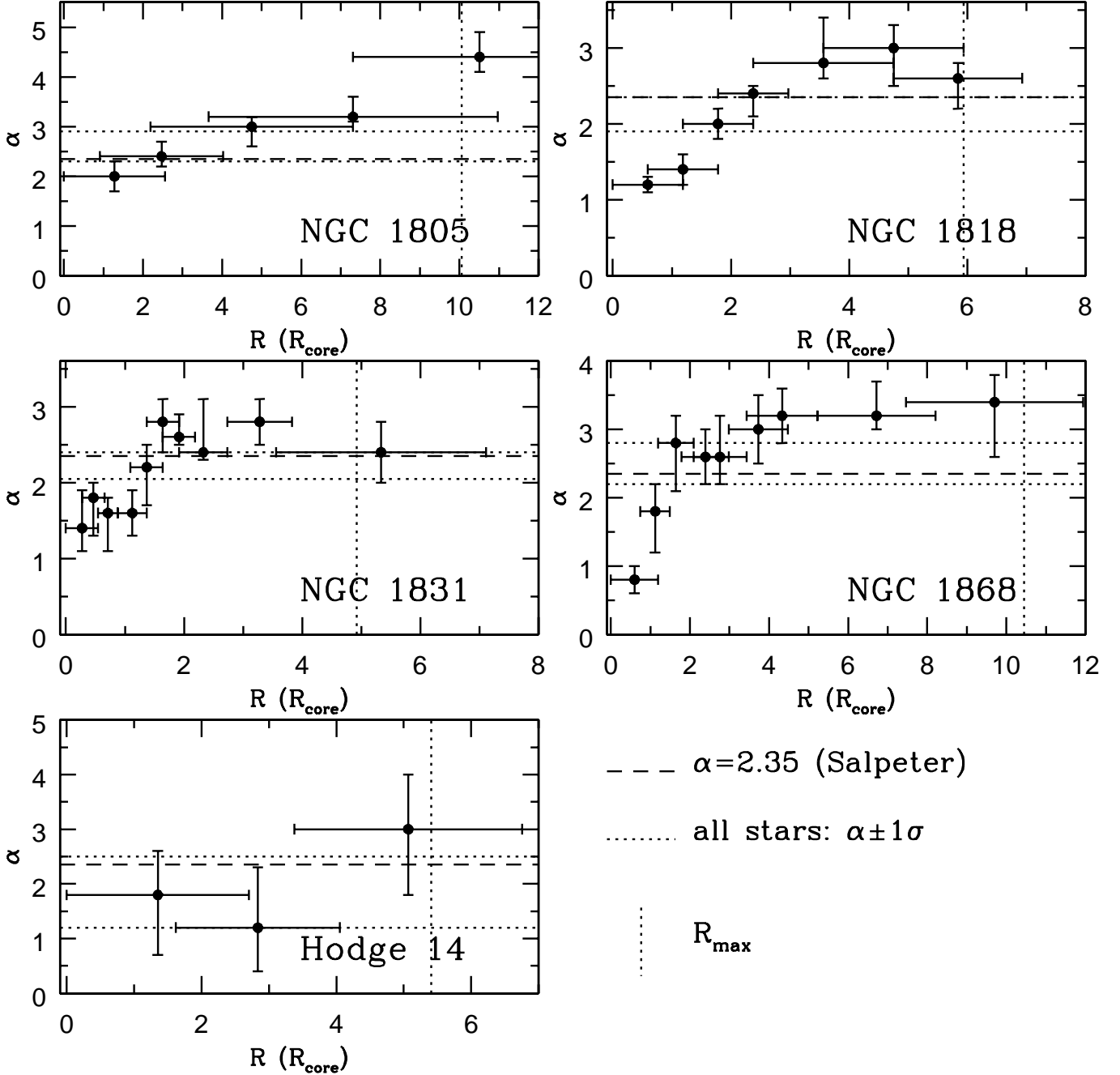}}
\caption[]{The PDMF slope vs. R in the 2D CMD analysis.
The lines are the same as in the Fig. \ref{alpha_R_1D}}
\label{alpha_R_2D}
\end{figure*}

Following this procedure for all clusters, we built Fig. 
\ref{alpha_R_2D}, the CMD method counterpart of Fig. \ref{alpha_R_1D}. 
Again all clusters show evidence of mass segregation. 
After taking the uncertainties into account, these results are in very good
agreement with those from the LF method. 

\section{Discussion}

As shown in the previous section the spatial dependence of $\alpha$ 
clearly reveals the presence of mass segregation in all clusters. 
Except in some cases where uncorrected for incompleteness effects
at lower masses may be strongly affecting the PDMF slope, the results 
in both LF and CMD methods used here are in clear agreement and
within the uncertainties. 

Previous, detailed spatial determinations of the PDMF slope were made only for 
NGC\,1805 (de Grijs et al. 2002a) and NGC\,1818 (de Grijs et al. 2002a, 
Gouliermis et al. 2005). 
In both cases these authors apply the LF method to their HST/WFPC2 
data to determine the $\Gamma$ (=-$\alpha$+1) behaviour with 
distance to the cluster centre. 
In general our results agree with theirs, even in 
the case where the data are restricted to a higher stellar mass range
($m \simgt 2.0 M_{\odot}$) (Gouliermis et al. 2005).
For the other clusters, there are no analyses of similar quality and
mass range in the literature. 

Notice that $\alpha \sim$ 2.35 in the interval $ 2 < R/R_{\rm{core}} < 3$ 
for all clusters, except for Hodge 14, for which $\alpha \sim$ 2.35
at $R/R_{\rm{core}} \simeq 4$. This range is close to the clusters' half-light
(or mass) radii.

\subsection{The unresolved binarity effect}
\label{bineffect}

\begin{figure}
\resizebox{\hsize}{!}{\includegraphics{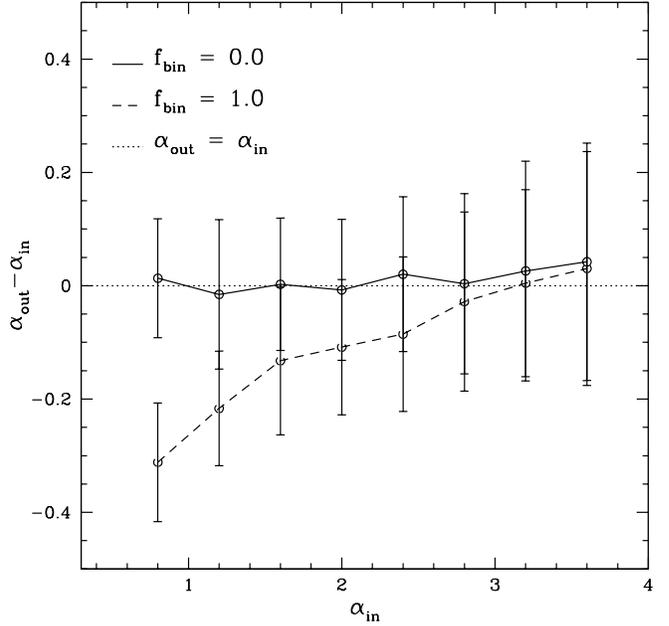}}
\caption[]{Control experiment to determine the effect of unresolved 
binaries in the PDMF slope recovered from a systemic LF. 
In the vertical axis we show the difference between the recovered slope 
$\alpha_{\rm{out}}$ and the input slope $\alpha_{\rm{in}}$.  
Solid line (dashed line): LF generated with $f_{\rm{bin}}=0\%$ 
($f_{\rm{bin}}=100\%$).}
\label{alphabin_1D}
\end{figure}

The major systematic effect over the usual LF method to determine the 
PDMF slope is the one caused by unresolved binaries; a flatter PDMF 
(lower $\alpha$) is likely to be recovered if binaries are not accounted for
(see the discussion in Sect. 3). 
In order to assess this issue directly in our approaches, 
we generate control experiments where we simulate artificial CMDs 
with $f_{\rm{bin}}=100\%$ and an input PDMF slope ($\alpha_{\rm{in}}$) 
and then recover the PDMF slope ($\alpha_{\rm{out}}$) by applying 
the LF method. 
As discussed in Sect. 3, although the primary and secondary stellar 
masses were drawn from the same PDMF,
only primary stars with mass $\simgt~0.9~M_{\odot}$ were directly taken 
into account, in accordance with the fainter magnitude limit in the
observed data.
Furthermore, random pairing and $\alpha=1.30$ for 
$0.08 \le m/M_{\odot} \le 0.8$ (unseen mass regime) yield  
$f_{\rm{bin,eff}}$ ($q \ge 0.60$) $\sim 0.20$.  
The results of these experiments are shown in Fig. \ref{alphabin_1D}.
As expected, the recovered PDMF is flatter than the input one, 
and the amplitude of this effect is greater for flatter input PDMFs. 
Except for $\alpha_{\rm{in}} < 1.2$, however the amplitude of 
$\alpha_{\rm{out}}-\alpha_{\rm{in}}$ is smaller than the uncertainties in 
$\alpha_{\rm{out}}$. 
Thus, for an unresolved binary fraction consistent with the one used in our 
CMD modelling process, its effect in the derived PDMF seems not to play 
a decisive role. 

\subsection{Dynamical mass segregation inside the core?}

To get some clues about the nature of the mass segregation, we 
evaluated the two-body relaxation time using the 
following expression (Binney \& Tremaine 1987):

$$ t_{rl} = \frac{6.5\rm{x}10^{8}}{\rm{ln}(0.4N)}
\left(\frac{M}{10^{5}M_{\odot}}\right)^{1/2}
\left(\frac{M_{\odot}}{m_{*}}\right)
\left(\frac{R}{pc}\right)^{3/2} $$

\noindent
where $N$ and $M$ are the total number of stars and the total mass
inside some radius $R$, respectively, and $m_{*}$ 
is a characteristic stellar mass. 
This is the time needed for stellar encounters to redistribute 
the stellar kinetic energy so that the 
the velocity distribution is approximately Maxwellian. 
However, it is important to keep in mind that 
this expression comes from analytical considerations.
The $t_{rl}$ obtained here allows a simple 
estimate of dynamical age for each cluster, since a 
more realistic treatment is only obtained through N-body simulations
(Kroupa, Aarseth, Hurley 2001; Baumgardt \& Makino 2003).

As discussed by Stolte et al. (2002), using the present physical conditions
of a cluster can only lead to the current dynamical status.
We therefore decided to evaluate $t_{rl}$ for {\it approximated initial
conditions}, where we take the IMF as that from Kroupa (2002) 
($\alpha=2.30$ for $0.50 < m/M_{\odot} \le 120$ and $\alpha=1.30$ for 
$0.08 < m \le 0.50 \Msun$) normalised
by the number of observed stars in a bright magnitude (massive) range. 
This range corresponds to $21.0 \le V_{555} \le 20.0$ 
($1.5 \simlt m/M_{\odot} \simlt 2.0$) for 
NGC\,1805, NGC\,1818, NGC\,1831, and NGC\,1868, and 
$22.0 \le V_{555} \le 21.0$ ($1.2 \simlt m/M_{\odot} \simlt 1.5$) 
for Hodge\,14. 
The characteristic stellar mass $m_{*}$  was taken to be the median mass 
for the adopted IMF inside the observed magnitude range, $m_{*}$ = 1.40 
$M_{\odot}$.
Table \ref{tab2} shows, for all clusters, the physical parameters 
estimated using this approach for all stars 
($0.08 \le m/M_{\odot} \le 120$) inside the core radius. 
The uncertainties in $N$ (Col. 3), $M$ (Col. 4), and $t_{rl}$ (Col. 5) 
result from the statistical fluctuations in the observed 
star counts ($N_{obs}$) (Col. 2) used in the IMF normalization. 

\begin{figure}
\resizebox{\hsize}{!}{\includegraphics{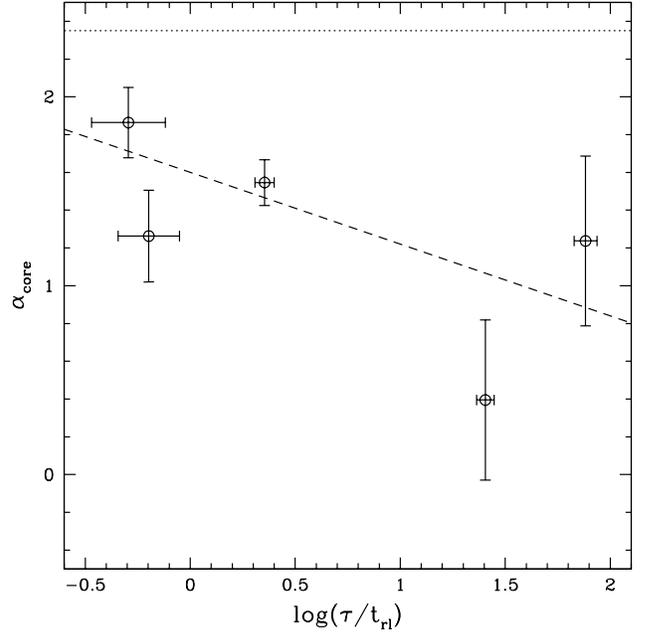}}
\caption[]{The PDMF slope vs. $\rm{log}(\tau/t_{rl})$ relation inside
the core radius.}
\label{alpha_trl_core}
\end{figure}

In this table we also present the ratio $\tau/t_{rl}$ in order to give an idea 
of the dynamical age of each cluster. The uncertainties in this ratio
were propagated from the corresponding uncertainties in $\tau$ and $t_{rl}$.
This ratio spans a wide range of values (0.2 to 30).
The relation between this parameter, evaluated for the core radius,
and the corresponding PDMF slope (determined only 
by the LF method due to low statistics) 
is plotted in Fig. \ref{alpha_trl_core}.
Although strongly based on the NGC\,1868 result, 
the relation suggests that dynamically younger clusters tend to have steeper 
PDMF than dynamically older ones, as expected from the effects of 
dynamical mass segregation.
We fit a linear relation (dashed line) between these parameters 
(correlation coefficient (c.c.)=0.64), with 
slope $\rm{d}\alpha_{core} / \rm{dlog}(\tau / t_{rl}) = -0.38 \pm 0.30$.
Also note that all PDMF are flatter than Salpeter
(marked as a horizontal dotted line in this figure).
By our fit one could expect that only clusters with log($\tau/t_{rl}$) 
$\sim -1.5$ would have a PDMF with a Salpeter slope within their cores. 

On the other hand, if NGC 1868 is omitted from the linear 
relation (c.c.=0.47), the slope becomes flatter ($-0.20 \pm 0.27$).
The lack of a strong trend in the central $\alpha$ value with age could
mean that mass segregation takes place at the onset of star formation
within a cluster (primordial segregation) or on such a short timescale that
it leaves little room for subsequent mass segregation due to dynamical 
effects on longer timescales. There is no specific work 
applying N-body simulations to address how fast the PMDF evolves in a 
typical globular cluster centre; in less dense environments, 
Kroupa, Aarseth, \& Hurley (2001) have shown
that the mass segregation in the ONC can be explained as a dynamical effect.
Notice also that there is a clear PDMF flattening with dynamical age 
inside the core radius of open clusters (Bonatto \& Bica 2005).

\begin{table}
\caption[]{Dynamical parameters and timescales estimated for the region inside 
the core radius.}
\label{tab2}
\small
\renewcommand{\tabcolsep}{0.8mm}
\begin{tabular}{lrrrrrr}
\hline\hline
Cluster	& $N_{obs}$ & $N$ & $M$ & $t_{rl}$ & $\tau/t_{rl}$ \cr
~~ & ~~ & $(\rm{x}10^{3})$ & $(\rm{x}10^{3}M_{\odot})$ & $(Myr)$ & \cr
\hline
NGC\,1805 & 66 & $3.9 \pm 0.5$ & $2.2 \pm 0.3$ & $16 \pm 4$ & $0.64 \pm 0.10$ \cr 
NGC\,1818 & 140 & $7.5 \pm 0.7$ & $4.1 \pm 0.4$ & $48 \pm 7$ & $0.51 \pm 0.10$ \cr
NGC\,1831 & 464 & $38.4 \pm 1.9$ & $20.9 \pm 1.2$ & $222 \pm 19$ & $2.26 \pm 0.10$ \cr  
NGC\,1868 & 182 & $16.5 \pm 1.3$ & $8.9 \pm 0.8$ & $35 \pm 5$ & $25.43 \pm 1.00$ \cr 
Hodge\,14 & 53 & $3.7 \pm 0.5$ & $2.1 \pm 0.3$ & $24 \pm 5$ & $76.25 \pm 4.00$ \cr
\hline
\end{tabular}
\end{table}

\subsection{Dynamical origin for the loss of lower-mass stars?}

\begin{figure}
\resizebox{\hsize}{!}{\includegraphics{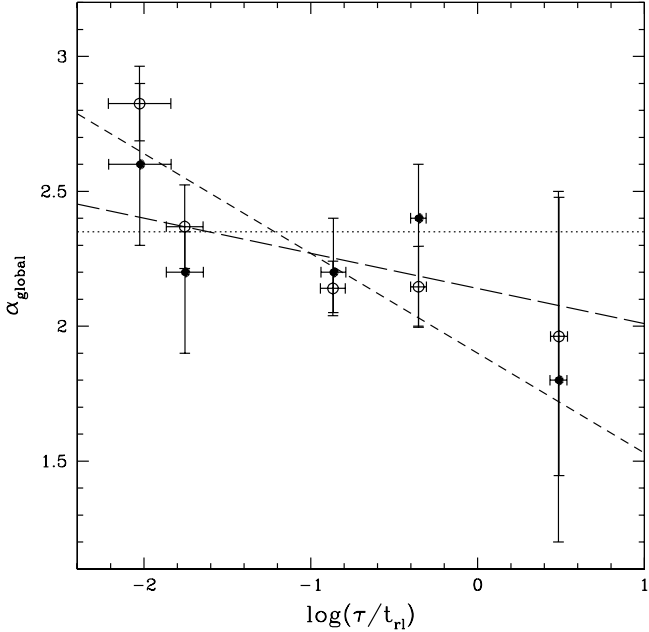}}
\caption[]{The PDMF slope vs. $\rm{log}(\tau/t_{rl})$ for the global cluster. 
Open (filled) circles correspond to LF (CMD) method. 
The best linear fit for the LF (CMD) method is shown as a
short(long)-dashed line. The dotted line corresponds to the Salpeter slope.}
\label{alpha_trl}
\end{figure}

In order to investigate a potential dynamical origin of the loss 
of lower-mass stars, we evaluated the same quantities as in the previous
section, but considering all stars in 
each cluster. The results are presented Table \ref{tab3}.
Although $t_{rl}$ does not correspond exactly to the
timescale for stellar evaporation, it 
is expected to scale with this one (Binney \& Tremaine 1987). 
As one can see in the last column of Table \ref{tab3}, the $\tau/t_{tl}$ 
values again span a wide range (4$\rm{x}$10$^{-3}$ to 1.2).
The relation between global $\alpha$ (from LF and CMD methods) and 
 global log($\tau/t_{tl}$) is plotted in Fig. \ref{alpha_trl}. 
For both methods there is a trend in the sense that dynamically older
clusters again have lower values of $\alpha$. 
Using the results of the LF method, we again fit a linear relation 
(short-dashed line, c.c.= 0.87) between these parameters and 
find a slope 
$\rm{d}\alpha_{\rm{global}} / \rm{dlog}(\tau / t_{rl}) = -0.37 \pm 0.12$. 
We also fit a linear relation using the results of the CMD method 
(long-dashed line, c.c.=0.50) and find a slope of $-0.13 \pm 0.13$.

It is very interesting to notice that this trend is consistent 
with recent results in the literature for a variety of systems. 
Using N-body simulations, Baumgardt \& Makino (2003) reproduce the 
observed shallowing in the global PDMF of globular clusters and
attribute this effect to the lower-mass stars preferentially depleted from
the cluster due to dynamical mass segregation. 
They also find that the details of this process are nearly independent 
of the starting conditions. 
Bonatto \& Bica (2005), analysing 2MASS data for 11 open clusters, 
clearly show that these systems also have the same effect. 
In both cases the MF flattening reaches at least 0.8 in slope, again 
consistent with what we observe in the rich LMC clusters.
Durgapal \& Pandey (2001) show that intermediate age and old open clusters
tend to have a smaller ratio between their present radius and their 
limiting radius as they become older. They attribute this to the
effect of loss of stars.
%However, the clusters analyzed by these authors do not present a clear 
%trend, perhapes because they are very dynamical old and similar.
%Furthermore, maybe it is not the best case to campare with LMC 
%clusters, since this lost of stars are probably very enhaced by 
%tidal forces due to the Milky Way disk.

\begin{table}
\caption[]{Estimates of the global dynamical parameters and 
timescales for the clusters.}
\label{tab3}
\small
\renewcommand{\tabcolsep}{0.8mm}
\begin{tabular}{lrrrrrr}
\hline\hline
Cluster	& $n_{obs}$ & $n_{tot}$ & $m_{tot}$ & $t_{rl}$ & $\tau/t_{rl}$ \cr
~~ & ~~ & $(\rm{x}10^{3})$ & $(\rm{x}10^{3}M_{\odot})$ & $(Gyr)$ & ~~ \cr
\hline
NGC\,1805 & 471 & $30.7 \pm 1.7$ & $16.8 \pm 0.9$ & $1.1 \pm 0.1$ & $0.010 \pm 0.002$ \cr 
NGC\,1818 & 774 & $46.3 \pm 2.1$ & $25.2 \pm 1.3$ & $1.4 \pm 0.1$ & $0.018 \pm 0.002$ \cr
NGC\,1831 & 1539 & $109.4 \pm 3.5$ & $59.5 \pm 2.1$ & $3.7 \pm 0.2$ & $0.135 \pm 0.010$ \cr  
NGC\,1868 & 929 & $63.9 \pm 1.3$ & $34.8 \pm 1.5$ & $2.0 \pm 0.1$ & $0.442 \pm 0.020$ \cr 
Hodge\,14 & 317 & $22.9 \pm 1.5$ & $12.5 \pm 0.8$ & $0.6 \pm 0.1$ & $3.073 \pm 0.150$ \cr
\hline
\end{tabular}
\end{table}

\section{Summary and conclusions}

We have analysed the shape of the PDMF at different radii for a
sample of 5 rich LMC clusters. The data are the result of deep imaging
with HST/WFPC2. 
The PDMFs were determined by two distinct approaches: 
1) a traditional method of converting stellar luminosities into masses; 
2) a full modelling of the colour-magnitude diagrams of the clusters, 
accounting for the effects of unresolved binaries and photometric errors.
Our results hold for both approaches and are insensitive to the 
details of how the PDMF and its slope are determined. 
Control experiments reveal that the unresolved binaries flatten the PDMF
recovered by the traditional methods. However this effect is less
than the uncertainties in the $\alpha$ determination.

We found significant mass segregation in all
of them. 
The effect is expected in the sense that the PDMF is steeper 
further out than in the core: 
$\alpha \simlt 1.80$ for $R \le 1$ R$_{\rm{core}}$ and
$\alpha \sim$ Salpeter (=2.35) inside $R=2 \sim 3$ R$_{\rm{core}}$
(except for Hodge 14, where it occurs at $R \sim$ 4R$_{\rm{core}}$).
Since the global PDMF is also near Salpeter, we confirm previous claims
that the PDMF evaluated around the half-mass radius is consistent 
with the global PDMF (Kroupa 2002).

The spatial dependence of the PDMF slope
was previously presented by de Grijs et al (2002a) 
for the youngest two clusters in our sample, 
NGC 1805 and NGC 1818, and by Gouliermis et al. (2005)
for NGC 1818.

We have investigated the origin of mass segregation 
and the loss of lower-mass stars by inferring a 
dynamical age ($\tau/t_{rl}$) for each cluster.
This was done in two regions: inside the cluster core and for the entire 
system. In both cases we notice that the dynamically older clusters
(with larger $\tau/t_{rl}$ values) tend to have shallower PDMFs.
Although this result is based in only 5 clusters, we interpret
this observed flattening trend as a dynamical effect, as also suggested by 
previous works for other systems 
(Kroupa, Aarseth \& Tout 2001; Baumgardt \& Makino 
2003; Bonatto \& Bica 2005). 
In fact, it is hard not to expect some dynamical
signature in the PDMFs for those LMC clusters with quite 
different dynamical ages.
On the other hand, the PDMF flattening 
(that can reach $\Delta \alpha \sim 1.0$) could be interpreted as 
primordial mass segregation, at least in the case of the central regions.
Or perhaps dynamical mass segregation in the 
LMC cluster's cores occurs fast enough that any age dependence of the PDMF
would be seen only in a sample of young clusters.
Again we stress the importance of N-body simulations
that configure a unique tool for providing reliable answers.

Although a scatter of $\sim 1.0$ in $\alpha$ is expected
due to Poisson noise and the dynamical evolution of stellar clusters
(Kroupa 2001), the aforementioned trends are in accordance with what one
expects from dynamical arguments. 
In order to improve this dynamical investigation, and therefore 
confirm or reject these trends, we are applying our PDMF determination
techniques to a larger number
of CMDs of LMC/SMC clusters (Brocato et al. 2001).

\begin{acknowledgements}
We thank Sandro Javiel and Charles Bonatto for useful discussions. 
We acknowledge CNPq and PRONEX/FINEP 76.97.1003.00 for partially 
supporting this work. LOK acknowledges FAPESP postdoctoral 
fellowship 05/01351-5.
\end{acknowledgements}

%sssssssssssssssssssssssssssss REFERENCESsssssssssssssssssssssssssssssss
%

\end{document}